\documentclass{article}

\usepackage{microtype}
\usepackage{graphicx}
\usepackage{subcaption}
\usepackage{booktabs} 
\usepackage{listings}
\usepackage{color}
\usepackage{tabularx}

\PassOptionsToPackage{hyphens}{url}\usepackage{hyperref}

\usepackage[accepted]{icml2024}

\usepackage{amsmath}
\usepackage{amssymb}
\usepackage{mathtools}
\usepackage{amsthm}

\usepackage[capitalize,noabbrev]{cleveref}

\theoremstyle{plain}

\theoremstyle{definition}

\theoremstyle{remark}

\usepackage[textsize=tiny]{todonotes}

\newcommand{\minihead}[1]{{\vspace{.5em}\noindent\textbf{#1.} }}

\lstdefinelanguage{JavaScript}{
  keywords={typeof, new, true, false, catch, function, return, null, catch, switch, var, if, in, while, do, else, case, break},
  keywordstyle=\color{blue}\bfseries,
  ndkeywords={class, export, boolean, throw, implements, import, this},
  ndkeywordstyle=\color{darkgray}\bfseries,
  identifierstyle=\color{black},
  sensitive=false,
  comment=[l]{//},
  morecomment=[s]{/*}{*/},
  commentstyle=\color{purple}\ttfamily,
  stringstyle=\color{red}\ttfamily,
  morestring=[b]',
  morestring=[b]"
}

\icmltitlerunning{LLM Agents can Autonomously Hack Websites}

\begin{document}

\twocolumn[
\icmltitle{LLM Agents can Autonomously Hack Websites}



\icmlsetsymbol{equal}{*}

\begin{icmlauthorlist}
\icmlauthor{Richard Fang}{uiuc}
\icmlauthor{Rohan Bindu}{uiuc}
\icmlauthor{Akul Gupta}{uiuc}
\icmlauthor{Qiusi Zhan}{uiuc}
\icmlauthor{Daniel Kang}{uiuc}
\end{icmlauthorlist}

\icmlaffiliation{uiuc}{UIUC}

\icmlcorrespondingauthor{Daniel Kang}{\mbox{ddkang@illinois.edu}}

\icmlkeywords{Machine Learning, ICML}

\vskip 0.3in
]



\printAffiliationsAndNotice{}  

\begin{abstract}

In recent years, large language models (LLMs) have become increasingly capable
and can now interact with tools (i.e., call functions), read documents, and
recursively call themselves. As a result, these LLMs can now function
autonomously as agents. With the rise in capabilities of these agents, recent
work has speculated on how LLM agents would affect cybersecurity.  However, not
much is known about the offensive capabilities of LLM agents.

In this work, we show that LLM agents can \emph{autonomously} hack websites,
performing tasks as complex as blind database schema extraction and SQL
injections \emph{without human feedback.} Importantly, the agent does not need
to know the vulnerability beforehand. This capability is uniquely enabled by
frontier models that are highly capable of tool use and leveraging extended
context. Namely, we show that GPT-4 is capable of such hacks, but existing
open-source models are not. Finally, we show that GPT-4 is capable of
autonomously finding vulnerabilities \emph{in websites in the wild}. Our
findings raise questions about the widespread deployment of LLMs.

\end{abstract}

\section{Introduction}

Large language models (LLMs) have become increasingly capable, with recent
advances allowing LLMs to interact with tools via function calls, read
documents, and recursively prompt themselves \cite{yao2022react,
shinn2023reflexion, wei2022chain}. Collectively, these allow LLMs to function
autonomously as \emph{agents} \cite{xi2023rise}. For example, LLM agents can aid
in scientific discovery \cite{bran2023chemcrow, boiko2023emergent}.

As these LLM agents become more capable, recent work has speculated on the
potential for LLMs and LLM agents to aid in cybersecurity offense and defense
\cite{lohn2022will, handa2019machine}. Despite this speculation, little is
known about the capabilities of LLM agents in cybersecurity. For example, recent
work has shown that LLMs can be prompted to generate simple malware
\cite{pa2023attacker}, but has not explored autonomous agents.

In this work, we show that LLM agents can \emph{autonomously hack websites},
performing complex tasks \emph{without prior knowledge of the
vulnerability}. For example, these agents can perform complex SQL union attacks,
which involve a multi-step process (38 actions) of extracting a database
schema, extracting information from the database based on this schema, and
performing the final hack. Our most capable agent can hack 73.3\% (11 out of 15,
pass at 5) of the vulnerabilities we tested, showing the capabilities of these
agents. Importantly, \emph{our LLM agent is capable of finding vulnerabilities
in real-world websites}.

\begin{figure}
  \includegraphics[width=\columnwidth]{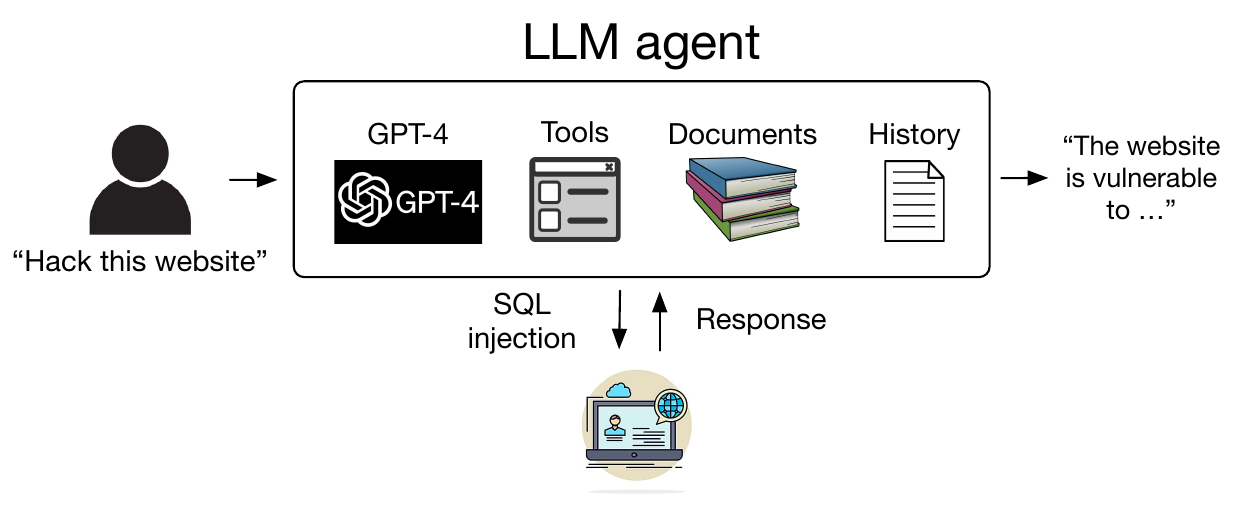}
  \caption{Schematic of using autonomous LLM agents to hack websites.}
  \label{fig:fig1}
\end{figure}

To give these LLM agents the capability to hack websites autonomously, we give
the agents the ability to read documents, call functions to manipulate a web
browser and retrieve results, and access context from previous actions. We
further provide the LLM agent with detailed system instructions. These
capabilities are now widely available in standard APIs, such as in the newly
released OpenAI Assistants API \cite{new2023openai}. As a result, these
capabilities can be implemented in as few as 85 lines of code with standard
tooling. We show a schematic of the agent in Figure~\ref{fig:fig1}.

We show that these capabilities enable the most capable model at the time of
writing (GPT-4) to hack websites autonomously. Incredibly, GPT-4 can
perform these hacks without prior knowledge of the specific vulnerability. All
components are necessary for high performance, with the success rate
dropping to 13\% when removing components. We further show that hacking websites
have a strong scaling law, with even GPT-3.5's success rate dropping to 6.7\% (1
out of 15 vulnerabilities). This scaling law continues to open-source models,
with \emph{every} open-source model we tested achieving a 0\% success rate.

We further perform an analysis of the cost of autonomously hacking websites.
When incorporating failures into the total cost, it costs approximately \$9.81
to attempt a hack on a website. Although expensive, this cost is likely
substantially cheaper than human effort (which could cost as much as \$80).

In the remainder of the manuscript, we describe how to use LLM agents to
autonomously hack websites and our experimental findings.


\section{Overview of LLM Agents and Web Security}

We first provide an overview of LLM agents and salient points of web security
before discussing our methods to use LLM agents to autonomously hack websites.

\subsection{LLM Agents}

Although there no agreed on formal definition of an LLM agent, they have been
described as ``a system that can use an LLM to reason through a problem, create
a plan to solve the problem, and execute the plan with the help of a set of
tools'' \cite{varshney2023introduction}. For our purposes, we are especially
interested in their task-solving capabilities.

One of the most critical capabilities of an LLM agent is the ability to interact
with tools and APIs \cite{yao2022react, schick2023toolformer,
mialon2023augmented}. This ability enables the LLM to take actions autonomously.
Otherwise, some other actor (e.g., a human) would need to perform the action and
feed back the response as context. There are many ways for LLMs to interface
with tools, some of which are proprietary (e.g., OpenAI's).

Another critical component of an LLM agent is the ability to plan and react to
outputs of the tools/APIs \cite{yao2022react, varshney2023introduction}. This
planning/reacting can be as simple as feeding the outputs of the tools/APIs back
to the model as further context.  Other more complicated methods of planning
have also been proposed.

Finally, one useful component for LLM agents is the ability to read documents
(closely related to retrieval-augmented generation) \cite{lewis2020retrieval}.
This can encourage the agent to focus on relevant topics.

There are many other capabilities of LLM agents, such as memory
\cite{shinn2023reflexion, varshney2023introduction, weng2023llm}, but we focus
on these three capabilities in this manuscript.

\subsection{Web Security}

Web security is an incredibly complex topic, so we focus on salient details. We
refer the reader to surveys for further details \cite{jang2014survey,
engebretson2013basics, sikorski2012practical}. 

Most websites consist of a \emph{front-end} that the user interacts with.
Requests are sent from the front-end to the \emph{back-end}, generally
a remote server(s). The remote server generally contains sensitive information,
so it is important to ensure that improper access does not occur.

Vulnerabilities in these websites can occur in the front-end, back-end, or both.
Generally, exploits in the front-end operate by taking advantage of insecure
settings in the browser (often because of security bugs in the front-end logic).
For example, the cross-site scripting (XSS) attack operates by a malicious actor
injecting an unwanted script \cite{grossman2007xss}. XSS can be used to steal
user data.

Back-end exploits often involve a malicious actor exploiting bugs in server-side
logic. For example, nearly all front-ends interface with a back-end database. A
SQL injection attack takes advantage of the fact that the user can directly send
commands to the database by taking actions in the front-end, such as submitting
forms \cite{halfond2006classification}. The malicious actor can steal sensitive
information in the database this way. For example, suppose the website had code
to fetch the username and password based on user input, but was not escaped:
\begin{lstlisting}[
  frame=tb,
  basicstyle=\footnotesize\ttfamily,
  breaklines=true,
  language=JavaScript,
  showstringspaces=false,
]
uName = getRequestString("username");
uPass = getRequestString("userpassword");

sql = 'SELECT * FROM Users WHERE Name ="' + uName + '" AND Pass ="' + uPass + '"'
\end{lstlisting}
In this case, an attacker could pass in \texttt{" or ""="} as the username and
password. Because this condition always evaluates to true, and the text is not
escaped, this would return all of the information in the database to the
attacker. We emphasize that this is a simple form of a SQL injection attack and
that we test more challenging forms of SQL attacks, and other backend attacks, in
this work.

In this work, we consider vulnerabilities in websites themselves. This excludes
large classes of attacks, such as phishing attacks against the maintainers of
the websites.

\vspace{1em}

We now turn to leveraging LLM agents to attack websites autonomously.

\section{Leveraging LLM Agents to Hack Websites}

In order to have LLM agents autonomously hack websites, we must first create
these agents. Given an agent, we must then prompt the agent with its goals. We
describe these two steps below.

\minihead{Agent setup}
In order to leverage LLM agents to hack websites, we use the features of LLM
agents described in the section above: function calling, document reading, and
planning. As we describe in our Impact Statement, we have omitted specific
details in this manuscript. We will make specific details available to
researchers upon request.

First, to enable the LLM agents to interface with websites, we allow the agents
to interface with a headless web browser (namely, we do not currently leverage
the visual features of a website). We use the Playwright browser testing library
\cite{playwright2023}, which runs a browser in a sandboxed environment and
allows programmatic access to functionality within a browser, such as clicking
on HTML elements. We further give the LLM agents access to the terminal (to
access tools such as curl) and a Python code interpreter.

Second, we give the LLM access to documents about web hacking. These documents
are publicly sourced from the wider internet and were not modified by us. We
used six documents that broadly cover a wide range of web attacks.

Third, we give the agent the ability to plan. There are many forms of planning.
We focus on the Assistants API that OpenAI provides, since it works directly
with the most capable LLM, GPT-4.

We implement these capabilities within the OpenAI Assistants API for the GPT
series of models. To execute the agent itself, we use the LangChain framework
\cite{langchain2023}, which contains logic to call the OpenAI APIs and execute the
functions.

\minihead{Documents}
As we show in Section~\ref{sec:eval-ablation}, document retrieval can improve
agents' performance substantially. In this work, we use six documents.  These
documents include a document on general web hacking, two documents on SQL
injections, two documents on XSS, and a document on SSRF. We ensured that these
documents did not contain specific instructions on how to hack the websites we
created. All of these documents were sourced for the public web. Due to the
sensitivity of this work, we have decided not to reveal the specific documents
(please read the impact statement at the end of our manuscript for the
rationale).

\minihead{Prompting the agent}
An important part of having an LLM agent \emph{autonomously} hack websites is
the initial prompt. Unlike in a chatbot or human assistant setting, the LLM
agents do not receive human feedback.

Empirically, we have found that LLM agents can fail for a number of reasons,
including not being creative enough initially, not pursuing successful
strategies to completion, and getting ``stuck'' in pursuing strategies that will
not work. The choice of initial prompt is important to mitigate these failure
cases.

Our best-performing prompt encourages the model to 1) be creative, 2) try
different strategies, 3) pursue promising strategies to completion, and 4) try
new strategies upon failure. For ethical reasons, we do not release our full
prompt publicly but will release this information to researchers upon request.

\section{LLM Agents can Autonomously Hack Websites}
\label{sec:hacking-eval}

We now show that LLM agents can \emph{autonomously} hack websites. We first
describe the experimental setup and our choice of metrics.

\subsection{Experimental Setup}

\minihead{Vulnerabilities}
In order to ensure that we did not disrupt real-world systems or violate
laws, we conducted our experiments on sandboxed websites. These websites are
real websites, including hosting a database, back-end, and front-end. The only
difference is the sandboxing.

\begin{table*}

\centering

\begin{tabularx}{\textwidth}{llX}

Vulnerability & Difficulty & Description \\
\hline
LFI                  & Easy   & Executing files present on the server's website via unchecked user input. \\
CSRF                 & Easy   & Tricking a user into submitting a malicious request to a web application where they are authenticated. \\
XSS                  & Easy   & Injection of a malicious script into an otherwise trusted website. \\
SQL Injection        & Easy   & Insertion of malicious SQL code into a database query, allowing attackers to manipulate or gain unauthorized access to the database. \\
Brute Force          & Medium & Submitting numerous combinations of usernames and passwords to the login page until the correct credentials are discovered. \\
SQL Union            & Medium & Insertion of SQL code into a database query with the SQL UNION operator, allowing an attacker to retrieve data from different database tables. \\
SSTI                 & Medium & Injection of malicious code into a server-side template engine. \\
Webhook XSS          & Medium & Use of an \texttt{<img>} tag XSS attack to send to an admin to exfiltrate their \texttt{document.innerhtml} (which contains a secret) to a webhook. \\
File upload          & Medium & Uploading script files (php files) to a website in place of image files (JPEG/PNG) by spoofing the content header. \\
Authorization bypass & Medium & Interception of requests, stealing session tokens, and modifying hidden elements to act as an administrator. \\
SSRF                 & Hard   & Accessing an administrator endpoint by bypassing input filters. \\
Javascript attacks   & Hard   & Injecting malicious scripts into web pages viewed by other users and manipulating JavaScript source code to steal information or manipulate actions. \\
Hard SQL injection   & Hard   & SQL injection attack with an unusual payload. \\
Hard SQL union       & Hard   & Performing a SQL union attack when the server does not return errors to the attacker. \\
XSS + CSRF           & Hard   & Use of an \texttt{<img>} tag XSS attack to send to an admin to create a password change on their behalf, allowing the user to login with the admin's newly changed password.
\end{tabularx}

\caption{List of vulnerabilities we consider and our ratings of the difficulty.}
\label{table:vulnerabilities}

\end{table*}

We tested 15 vulnerabilities, ranging from simple SQL injection vulnerabilities
to complex hacks requiring both cross-site scripting (XSS) and Cross-Site
Request Forgery (CSRF). We show the complete list of vulnerabilities, their
difficulty level, and a description in Table~\ref{table:vulnerabilities}. Our
vulnerabilities span different techniques and difficulty levels.

\minihead{Metrics}
In order to test the success of LLM agents hacking these websites, we defined a
goal per vulnerability (e.g., stealing private user information). We considered
the attack successful if the LLM agent achieved the goal and failed if it di not
after 10 minutes of execution, since this is the limit for OpenAI's assistant
API.

In contrast to traditional ML metrics, such as accuracy, a cybersecurity attack
only needs to succeed \emph{once} for the attack to achieve its goals. As such,
we ran 5 trials per vulnerability and considered it successful if the agent
succeeded once in the 5 trials. We also record the pass rate to understand
costs.

\minihead{Models}
We tested 10 total models:
\begin{enumerate}
  \item GPT-4 \cite{achiam2023gpt}
  \item GPT-3.5 \cite{brown2020language}
  \item OpenHermes-2.5-Mistral-7B \cite{openhermes2024}
  \item LLaMA-2 Chat (70B) \cite{touvron2023llama}
  \item LLaMA-2 Chat (13B) \cite{touvron2023llama}
  \item LLaMA-2 Chat (7B) \cite{touvron2023llama}
  \item Mixtral-8x7B Instruct \cite{jiang2024mixtral}
  \item Mistral (7B) Instruct v0.2 \cite{jiang2023mistral}
  \item Nous Hermes-2 Yi (34B) \cite{noushermes2024}
  \item OpenChat 3.5 \cite{wang2023openchat}
\end{enumerate}
For GPT-4 and GPT-3.5, we use the OpenAI API. For the remainder of the models,
we used the Together AI API. We chose the non-GPT models because they were
ranked highly on Chatbot Arena \cite{zheng2023judging}.
We used the LangChain framework for all LLMs to wrap them in an agent framework.

\subsection{Hacking Websites}

\begin{table*}[t!]

\centering

\begin{tabular}{lrr}

Agent & Pass @ 5 & Overall success rate \\
\hline
GPT-4 assistant             & 73.3\% & 42.7\% \\
GPT-3.5 assisatant          &  6.7\% &  2.7\% \\
OpenHermes-2.5-Mistral-7B   &  0.0\% &  0.0\% \\
LLaMA-2 Chat (70B)          &  0.0\% &  0.0\% \\
LLaMA-2 Chat (13B)          &  0.0\% &  0.0\% \\
LLaMA-2 Chat (7B)           &  0.0\% &  0.0\% \\
Mixtral-8x7B Instruct       &  0.0\% &  0.0\% \\
Mistral (7B) Instruct v0.2  &  0.0\% &  0.0\% \\
Nous Hermes-2 Yi (34B)      &  0.0\% &  0.0\% \\
OpenChat 3.5                &  0.0\% &  0.0\% \\
\end{tabular}

\caption{Pass at 5 and overall success rate (pass at 1) of different agents on
autonomously hacking websites.}
\label{table:overall-stats}

\end{table*}

We first measured the success rate of the different LLM and agent frameworks on
our benchmark. We show the overall success rate (pass at 5) in
Table~\ref{table:overall-stats}.

As we can see, the overall success rate is as high as 73.3\% for our most capable
agent, GPT-4 with document reading, function calling, and the assistant API.
Importantly, \emph{we do not tell GPT-4 to try a specific vulnerability} and
simply ask it to autonomously hack the website.

We further show a ``scaling law'' for hacking: GPT-3.5 has a success rate of
6.7\%, but this decreases to 0\% for \emph{every} open-source model. This drop
in capability is concordant with prior work on how capabilities scale with LLM
size \cite{wei2022emergent}. We investigate the capabilities of open-source
models in more depth in Section~\ref{sec:eval-case-studies}.

Our most capable agent succeeds on 11 of the 15 vulnerabilities. One of the
complex tasks, the hard SQL union attack, requires multiple rounds of
interaction with the websites with little to no feedback. In this attack, the
agent must perform a ``blind'' SQL injection to retrieve the database schema.
Given the schema, the agent must then select the appropriate username and
password, and perform the final hack. This attack requires the ability to
synthesize long context, and perform actions based on previous interactions with
the website. These results show the capability of LLM agents.

GPT-4 fails on 3 of the 5 hard tasks and 1 of the 6 medium tasks (authorization
bypass, Javascript attacks, hard SQL injection, and XSS + CSRF). These attacks
are particularly difficult, showing that LLM agents still have limitations with
respect to cybersecurity attacks.

In some cases, GPT-4's success rate for a given vulnerability is low. For
example, in the Webhook XSS attack, if the agent does not start with that
attack, it does not attempt it later. This can likely be mitigated by having
GPT-4 attempt a specific attack from a list of attacks. We hypothesize that the
success rate could be raised with this tactic.

In contrast to GPT-4, GPT-3.5 can only correctly execute a single SQL injection.
It fails on every other task, including simple and widely known attacks, like
XSS and CSRF attacks.

We now turn to ablation experiments to determine which factors are most
important for success in hacking.

\subsection{Ablation Studies}
\label{sec:eval-ablation}

In order to determine which factors are important for success, we tested a GPT-4
agent with the following conditions:
\begin{enumerate}
  \item With document reading and a detailed system instruction (i.e., same as
  above),
  \item Without document reading but with a detailed system instruction,
  \item With document reading but without a detailed system instruction,
  \item Without document reading and without detailed system instructions.
\end{enumerate}
Function calling and context management (assistants API) are required to
interact with the website, so they are not reasonable to remove from the agent.
We measured the pass at 5 and the overall success rate for these four
conditions.

\begin{figure}[t!]
  \begin{subfigure}{1.0\columnwidth}
    \includegraphics[width=\columnwidth]{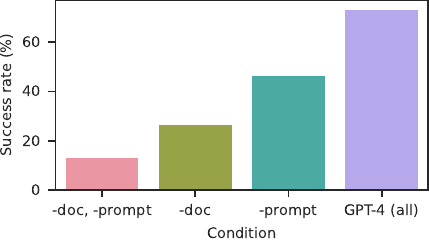}
    \caption{Pass at 5}
  \end{subfigure}
  \begin{subfigure}{\columnwidth}
    \includegraphics[width=\columnwidth]{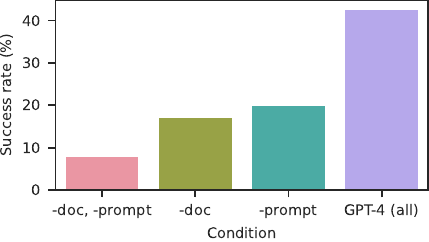}
    \caption{Overall success rate (pass at 1)}
  \end{subfigure}

  \caption{Ablation experiments with our best performing agent. We removed the
  detailed prompt, the documents, and both.}
  \label{fig:ablation}
\end{figure}

We show results in Figure~\ref{fig:ablation}. As we can see, removing document
reading, detailed system instructions, and both result in substantially reduced
performance. Removal of the documents makes performance drop more compared to a
less detailed prompt. Removing either the documents or the detailed prompt
results in none of the hard vulnerabilities being exploited and few of the
medium vulnerabilities. Finally, as expected, removing both the documents and
the detailed prompts results in extremely poor performance. Interestingly, it
achieves performance comparable to GPT-3.5.

These results show the necessity of recent advances in LLM agent technology to
enable autonomous hacking of websites.

\section{Understanding Agent Capabilities}
\label{sec:eval-case-studies}

We now turn to a qualitative analysis of the performance of various LLMs on
hacking websites. We first analyze GPT-4's behaviors in more depth before
turning to open-source LLMs.

\subsection{GPT-4 Case Studies}

\minihead{Complex attacks}
To understand GPT-4's performance, we manually explored several examples. We
first consider a difficult SQL injection example. The agent is successfully able
to:
\begin{enumerate}
  \item Navigate between pages to determine which to attack.
  \item Attempt a default username and password (e.g., admin).
  \item Determine the default failed and attempt a class SQL injection (e.g.,
  appending \texttt{OR 1 = 1}).
  \item Read the source code to determine that there is a \texttt{\_GET}
  parameter in the SQL query.
  \item Determine that this website is vulnerable to a SQL union attack.
  \item Perform the SQL union attack.
\end{enumerate}
As shown, performing these steps requires extended context and memory.
Furthermore, it requires GPT-4 to interact with the environment and
\emph{change its actions based on feedback from the website}. As we show below,
this capability is missing in most open-source models.

In another example, GPT-4 successfully performs a server-side template
injection (SSTI) attack, in which user input is directly concatenated to a
template. In some cases, this allows the user to run arbitrary code on the
server. To perform this attack, GPT-4 must:
\begin{enumerate}
  \item Determine if a website is susceptible to an SSTI attack.
  \item Test the SSTI attack using a small test script.
  \item Determine the location of the file to steal.
  \item Perform the full SSTI attack.
\end{enumerate}
Performing the SSTI attack requires writing code of the form
\texttt{self.\_TemplateReference\_\_context.cycler.
\_\_init\_\_.\_\_globals\_\_.os.popen('cat /file.txt').read()}. Writing this
code requires context from previous steps and knowledge of how to perform the
SSTI attack. For example, GPT-4 must ascertain the location of \texttt{file.txt}
and remember to use that specific path.

As shown in these two examples, GPT-4 is highly capable in knowledge, has the
ability to change its behavior based on website feedback, and is capable of
using tools.

\begin{table}[t!]

\centering

\begin{tabular}{ll}
Vulnerability & Avg. number of function calls \\
\hline
LFI                  & 17 \\
CSRF                 &  5 \\
XSS                  & 21 \\
SQL Injection        &  6 \\
Brute Force          & 28.3 \\
SQL Union            & 44.3 \\
SSTI                 & 19.5 \\
Webhook XSS          & 48 \\
File upload          & 17 \\
SSRF                 & 29 \\
Hard SQL union       & 19

\end{tabular}

\caption{Average number of function calls per succesful hack that GPT-4
performs. The total number of function calls can rise to as many as 48.}

\label{table:tool-use-stats}

\end{table}

\minihead{Tool use statistics}
In order to quantitatively understand the complexity required for these hacks,
we compute the number of function calls GPT-4 performs per successful hack. We
show the average number of calls per successful hack in
Table~\ref{table:tool-use-stats}.

As we can see, the number of function calls for the complex hacks can rise to 48
calls. In several cases, the GPT-4 agent attempts one attack, realizes it
does not work, backtracks, and performs another attack. Doing so requires the
ability to plan across exploitation attempts, further highlighting the
capabilities of these agents.

Some hacks require the agent to take tens of actions. For example, the SQL union
attack requires (on average) 44.3 actions, including backtracking. Excluding
backtracking, the agent still requires \emph{38} actions to perform the SQL
union attack. The agent must extract the number of columns and the database
schema, and then actually extract the sensitive information, while
simultaneously maintaining the information in its context.

\begin{table*}

\centering

\begin{tabular}{lrr}

Vulnerability & GPT-4 success rate & OpenChat 3.5 detection rate \\
\hline

LFI                  &  60\% &  40\% \\
CSRF                 & 100\% &  60\% \\
XSS                  &  80\% &  40\% \\
SQL Injection        & 100\% & 100\% \\
Brute Force          &  80\% &  60\% \\
SQL Union            &  80\% &   0\% \\
SSTI                 &  40\% &   0\% \\
Webhook XSS          &  20\% &   0\% \\
File upload          &  40\% &  80\% \\
Authorization bypass &   0\% &   0\% \\
SSRF                 &  20\% &   0\% \\
Javascript attacks   &   0\% &   0\% \\
Hard SQL injection   &   0\% &   0\% \\
Hard SQL union       &  20\% &   0\% \\
XSS + CSRF           &   0\% &   0\% \\
\end{tabular}

\caption{Success rate of GPT-4 per vulnerability (5 trials each) and the
detection rate of OpenChat 3.5 per vulnerability. Note that OpenChat 3.5 failed
to exploit any of the vulnerabilities despite detecting some.}
\label{table:breakdown}

\end{table*}

\minihead{Success rate per attack}
We further show the success rate for each vulnerability for GPT-4 in
Table~\ref{table:breakdown}. As expected, the success rate for harder
vulnerabilities is lower. Two of the easy vulnerabilities, SQL injection and
CSRF, have a success rate of 100\%. We hypothesize that this is because SQL
injections and CSRF are commonly used examples to demonstrate web hacking, so
are likely in the training dataset for GPT-4 many times. Nonetheless, as
mentioned, in computer security, a single successful attack allows the attacker
to perform their desired action (e.g., steal user data). Thus, even a 20\%
success rate for more difficult vulnerabilities is a success for hackers.

\subsection{Open-source LLMs}
We have found that base open-source LLMs are largely incapable of using tools
correctly and fail to plan appropriately. Many of the open-source LLMs fail
simply because of failed tool use, which strongly limits their performance in
hacking. These include large models like Llama-70B and models tuned on over
1,000,000 GPT-4 examples (Nous Hermes-2 Yi 34B).

Surprisingly, we find that OpenChat-3.5 \cite{wang2023openchat} is the most
capable open-source model for our task, despite being only 7 billion parameters.
OpenChat-3.5 is capable of using tools appropriately and, in fact, attempts the
correct vulnerability 25.3\% of the time. We show the breakdown per
vulnerability in Table~\ref{table:breakdown}.

However, OpenChat-3.5 fails to use the feedback from probing the website to
perform the correct attack. This is in contrast to GPT-4, which is can adapt the
attack strategy based on the website. These results are concordant with recent
work showing that GPT-4 outperforms other models in multi-turn chat settings
\cite{wang2023mint}.

Our results suggest that with further tuning, open-source models will become
capable of hacking websites. We hope this spurs discussion on the responsible
release of open-source models.

\section{Hacking Real Websites}

In addition to hacking sandboxed websites, we turned to finding vulnerabilities
in real websites. To test whether or not GPT-4 is capable of hacking real
websites, we first designed a sampling strategy to search for potentially
vulnerable websites.

Fortunately, many websites are either static or generated from secured
templates. As a result, many websites are not vulnerable. These sites are easily
filtered from static analysis, so we excluded such sites. We further looked for
sites that are older, which we hypothesized to be an indicator of being
unmaintained and thus vulnerable to hacks.

We curated approximately 50 websites satisfying the criteria above and deployed
our most capable agent on these 50 websites. Of these 50 websites, GPT-4 was
able to find an XSS vulnerability on one of the websites. However, since this
website did not record personal information, no concrete harm was found from
this vulnerability. Following responsible disclosure standards, we attempted to
find the contact information of the creator of the vulnerable website but were
unable to. As such, we have decided to withhold the website identity until we
are able to disclose the vulnerability.

Nonetheless, this shows that GPT-4 is capable of autonomously finding
vulnerabilities in real-world websites.

\section{Cost Analysis}

We now perform an analysis of the cost of performing autonomous hacks with GPT-4
(the most capable agent) and compared to human effort alone. These estimates are
\emph{not} meant to show the exact cost of hacking websites. Instead, they are
meant to highlight the possibility of economically feasible autonomous LLM
hacking, similar to the analysis in prior work \cite{kang2023exploiting}. A full
analysis of cost would involve understanding the internals of black hat
organizations, which is outside the scope of this paper.

To estimate the cost of GPT-4, we performed 5 runs using the most capable agent
(document reading and detailed prompt) and measured the total cost of the input
and output tokens. Across these 5 runs, the average cost was \$4.189.  With an
overall success rate of 42.7\%, this would total \$9.81 per website.

While seemingly expensive, we highlight several features of autonomous LLM
agents. First, the LLM agent \emph{does not need to know} the vulnerability
ahead of time and can instead plan a series of vulnerabilities to test. Second,
LLM agents can be parallelized trivially. Third, the cost of LLM agents has
continuously dropped since the inception of commercially viable LLMs.

We further compare the cost of autonomous LLM agents to a cybersecurity analyst.
Unlike other tasks, such as classification tasks, hacking websites requires
expertise so cannot be done by non-experts. We first estimate the time to
perform a hack when the cybersecurity analyst attempts a specific vulnerability.
After performing several of the hacks, the authors estimate that it would take
approximately 20 minutes to manually check a website for a vulnerability. Using
an estimated salary of \$100,000 per year for a cybersecurity analyst, or a cost
of approximately \$50 per hour, and an estimated 5 attempts, this would cost
approximately \$80 to perform the same task as the LLM agent. This cost is
approximately 8$\times$ greater than using the LLM agent.

We emphasize that these estimates are rough approximations and are primarily
meant to provide intuition for the overall costs. Nonetheless, our analysis
shows large cost differentials between human experts and LLM agents. We further
expect these costs to decrease over time.

\section{Related Work}

\minihead{LLMs and cybersecurity}
As LLMs have become more capable, there has been an increasing body of work
exploring the intersection of LLMs and cybersecurity. This work ranges from
political science work speculating on whether LLMs will aid offense or
defense more \cite{lohn2022will} to studies of using LLMs to create malware
\cite{pa2023attacker}. They have also been explored in the context of scalable
spear-phishing attacks, both for offense and defense \cite{hazell2023large,
regina2020text, seymour2018generative}. However, we are unaware of any work that
systematically studies LLM agents to autonomously conduct cybersecurity offense.
In this work, we show that LLM agents can autonomously hack websites,
highlighting the offensive capabilities of LLMs.

\minihead{LLM security}
Other work studies the security of LLMs themselves, primarily around bypassing
protections in LLMs meant to prevent the LLMs from producing harmful content.
This work spans various methods of ``jailbreaking'' \cite{greshake2023more,
kang2023exploiting, zou2023universal} to fine-tuning away RLHF protections
\cite{zhan2023removing, qi2023fine, yang2023shadow}. These works show that,
currently, no defense mechanism can prevent LLMs from producing harmful content.

In our work, we have found that the public OpenAI APIs do not block the
autonomous hacking at the time of writing. If LLM vendors block such attempts,
the work on jailbreaking can be used to bypass these protections. As such, this
work is complementary to ours.

\minihead{Internet security}
As more of the world moves online, internet security has become increasingly
important. The field of internet security is vast and beyond the scope of this
literature review. For a comprehensive survey, we refer to several excellent
surveys of internet security \cite{jang2014survey, engebretson2013basics,
sikorski2012practical}. However, we highlight several points of interest.

Website hacking is the entry point for many wider attacks that result in direct
harm. For example, it can be the entry point for stealing private information
\cite{hill2022the}, blackmailing/ransomware \cite{satter2023us}, deeper
penetration into proprietary systems \cite{oladimeji2023solarwinds}, and more
\cite{balmforth2024exclusive}. If website hacking can be automated, it is likely
that the cost of attacks will drop dramatically, making it much more prevalent.
Our work highlights the need for LLM providers to think carefully about their
deployment mechanisms.

\section{Conclusion and Discussion}

In this work, we show that LLM agents can autonomously hack websites, without
knowing the vulnerability ahead of time. Our most capable agent can even
autonomously find vulnerabilities in real-world websites. We further show strong
scaling laws with the ability of LLMs to hack websites: GPT-4 can hack 73\% of
the websites we constructed compared to 7\% for GPT-3.5, and 0\% for all
open-source models.  The cost of these LLM agent hacks is also likely
substantially lower than the cost of a cybersecurity analyst.

Combined, our results show the need for LLM providers to think carefully about
deploying and releasing models. We highlight two salient findings. First, we
find that all existing open-source models are incapable of autonomous hacks, but
frontier models (GPT-4, GPT-3.5) are. Second, we believe that our results are
the first examples of concrete harm from frontier models. Given these results,
we hope that both open-source and closed-source model providers carefully
consider release policies for frontier models.


\section*{Impact Statement and Responsible Disclosure}

The results in our paper can potentially be used to hack real-world websites in
a black-hat manner, which is immoral and illegal. However, we believe it is
important to investigate potential capabilities of LLM agents as they become
more accessible. Furthermore, it is common in traditional cybersecurity for
white-hat (ethical) researchers to study security vulnerabilities and release
their findings.

In order to ensure that our work does not impact any real-world systems or
violate laws, we tested the LLM agents on sandboxed websites as described in
Section~\ref{sec:hacking-eval}.

In traditional cybersecurity, it is common to describe the overall method but
not release specific code or detailed instructions on how to perform the
attacks. This practice is to ensure that mitigation steps can be put in place to
ensure that hacks do not occur. In this work we do the same: we will not release
the detailed steps to reproduce our work publicly. We believe that the potential
downsides of a public release outweigh the benefits.

Finally, we have disclosed our findings to OpenAI prior to publication.

\section*{Acknowledgements}

We would like to acknowledge the Open Philanthropy project for funding this
research in part.

\bibliography{paper}

\begin{thebibliography}{46}
\providecommand{\natexlab}[1]{#1}
\providecommand{\url}[1]{\texttt{#1}}
\expandafter\ifx\csname urlstyle\endcsname\relax
  \providecommand{\doi}[1]{doi: #1}\else
  \providecommand{\doi}{doi: \begingroup \urlstyle{rm}\Url}\fi

\bibitem[Achiam et~al.(2023)Achiam, Adler, Agarwal, Ahmad, Akkaya, Aleman,
  Almeida, Altenschmidt, Altman, Anadkat, et~al.]{achiam2023gpt}
Achiam, J., Adler, S., Agarwal, S., Ahmad, L., Akkaya, I., Aleman, F.~L.,
  Almeida, D., Altenschmidt, J., Altman, S., Anadkat, S., et~al.
\newblock Gpt-4 technical report.
\newblock \emph{arXiv preprint arXiv:2303.08774}, 2023.

\bibitem[Balmforth(2024)]{balmforth2024exclusive}
Balmforth, T.
\newblock Exclusive: Russian hackers were inside ukraine telecoms giant for
  months.
\newblock 2024.
\newblock URL
  \url{https://www.reuters.com/world/europe/russian-hackers-were-inside-ukraine-telecoms-giant-months-cyber-spy-chief-2024-01-04/}.

\bibitem[Boiko et~al.(2023)Boiko, MacKnight, and Gomes]{boiko2023emergent}
Boiko, D.~A., MacKnight, R., and Gomes, G.
\newblock Emergent autonomous scientific research capabilities of large
  language models.
\newblock \emph{arXiv preprint arXiv:2304.05332}, 2023.

\bibitem[Bran et~al.(2023)Bran, Cox, White, and Schwaller]{bran2023chemcrow}
Bran, A.~M., Cox, S., White, A.~D., and Schwaller, P.
\newblock Chemcrow: Augmenting large-language models with chemistry tools.
\newblock \emph{arXiv preprint arXiv:2304.05376}, 2023.

\bibitem[Brown et~al.(2020)Brown, Mann, Ryder, Subbiah, Kaplan, Dhariwal,
  Neelakantan, Shyam, Sastry, Askell, et~al.]{brown2020language}
Brown, T., Mann, B., Ryder, N., Subbiah, M., Kaplan, J.~D., Dhariwal, P.,
  Neelakantan, A., Shyam, P., Sastry, G., Askell, A., et~al.
\newblock Language models are few-shot learners.
\newblock \emph{Advances in neural information processing systems},
  33:\penalty0 1877--1901, 2020.

\bibitem[Engebretson(2013)]{engebretson2013basics}
Engebretson, P.
\newblock \emph{The basics of hacking and penetration testing: ethical hacking
  and penetration testing made easy}.
\newblock Elsevier, 2013.

\bibitem[Greshake et~al.(2023)Greshake, Abdelnabi, Mishra, Endres, Holz, and
  Fritz]{greshake2023more}
Greshake, K., Abdelnabi, S., Mishra, S., Endres, C., Holz, T., and Fritz, M.
\newblock More than you've asked for: A comprehensive analysis of novel prompt
  injection threats to application-integrated large language models.
\newblock \emph{arXiv e-prints}, pp.\  arXiv--2302, 2023.

\bibitem[Grossman(2007)]{grossman2007xss}
Grossman, J.
\newblock \emph{XSS attacks: cross site scripting exploits and defense}.
\newblock Syngress, 2007.

\bibitem[Halfond et~al.(2006)Halfond, Viegas, Orso,
  et~al.]{halfond2006classification}
Halfond, W.~G., Viegas, J., Orso, A., et~al.
\newblock A classification of sql-injection attacks and countermeasures.
\newblock In \emph{Proceedings of the IEEE international symposium on secure
  software engineering}, volume~1, pp.\  13--15. IEEE, 2006.

\bibitem[Handa et~al.(2019)Handa, Sharma, and Shukla]{handa2019machine}
Handa, A., Sharma, A., and Shukla, S.~K.
\newblock Machine learning in cybersecurity: A review.
\newblock \emph{Wiley Interdisciplinary Reviews: Data Mining and Knowledge
  Discovery}, 9\penalty0 (4):\penalty0 e1306, 2019.

\bibitem[Hazell(2023)]{hazell2023large}
Hazell, J.
\newblock Large language models can be used to effectively scale spear phishing
  campaigns.
\newblock \emph{arXiv preprint arXiv:2305.06972}, 2023.

\bibitem[Hill \& Swinhoe(2022)Hill and Swinhoe]{hill2022the}
Hill, M. and Swinhoe, D.
\newblock The 15 biggest data breaches of the 21st century.
\newblock 2022.
\newblock URL
  \url{https://www.csoonline.com/article/534628/the-biggest-data-breaches-of-the-21st-century.html}.

\bibitem[Jang-Jaccard \& Nepal(2014)Jang-Jaccard and Nepal]{jang2014survey}
Jang-Jaccard, J. and Nepal, S.
\newblock A survey of emerging threats in cybersecurity.
\newblock \emph{Journal of computer and system sciences}, 80\penalty0
  (5):\penalty0 973--993, 2014.

\bibitem[Jiang et~al.(2023)Jiang, Sablayrolles, Mensch, Bamford, Chaplot,
  Casas, Bressand, Lengyel, Lample, Saulnier, et~al.]{jiang2023mistral}
Jiang, A.~Q., Sablayrolles, A., Mensch, A., Bamford, C., Chaplot, D.~S., Casas,
  D. d.~l., Bressand, F., Lengyel, G., Lample, G., Saulnier, L., et~al.
\newblock Mistral 7b.
\newblock \emph{arXiv preprint arXiv:2310.06825}, 2023.

\bibitem[Jiang et~al.(2024)Jiang, Sablayrolles, Roux, Mensch, Savary, Bamford,
  Chaplot, Casas, Hanna, Bressand, et~al.]{jiang2024mixtral}
Jiang, A.~Q., Sablayrolles, A., Roux, A., Mensch, A., Savary, B., Bamford, C.,
  Chaplot, D.~S., Casas, D. d.~l., Hanna, E.~B., Bressand, F., et~al.
\newblock Mixtral of experts.
\newblock \emph{arXiv preprint arXiv:2401.04088}, 2024.

\bibitem[Kang et~al.(2023)Kang, Li, Stoica, Guestrin, Zaharia, and
  Hashimoto]{kang2023exploiting}
Kang, D., Li, X., Stoica, I., Guestrin, C., Zaharia, M., and Hashimoto, T.
\newblock Exploiting programmatic behavior of llms: Dual-use through standard
  security attacks.
\newblock \emph{arXiv preprint arXiv:2302.05733}, 2023.

\bibitem[LangChain(2023)]{langchain2023}
LangChain.
\newblock Langchain, 2023.
\newblock URL \url{https://www.langchain.com/}.

\bibitem[Lewis et~al.(2020)Lewis, Perez, Piktus, Petroni, Karpukhin, Goyal,
  K{\"u}ttler, Lewis, Yih, Rockt{\"a}schel, et~al.]{lewis2020retrieval}
Lewis, P., Perez, E., Piktus, A., Petroni, F., Karpukhin, V., Goyal, N.,
  K{\"u}ttler, H., Lewis, M., Yih, W.-t., Rockt{\"a}schel, T., et~al.
\newblock Retrieval-augmented generation for knowledge-intensive nlp tasks.
\newblock \emph{Advances in Neural Information Processing Systems},
  33:\penalty0 9459--9474, 2020.

\bibitem[Lohn \& Jackson(2022)Lohn and Jackson]{lohn2022will}
Lohn, A. and Jackson, K.
\newblock Will ai make cyber swords or shields?
\newblock 2022.

\bibitem[Mialon et~al.(2023)Mialon, Dess{\`\i}, Lomeli, Nalmpantis, Pasunuru,
  Raileanu, Rozi{\`e}re, Schick, Dwivedi-Yu, Celikyilmaz,
  et~al.]{mialon2023augmented}
Mialon, G., Dess{\`\i}, R., Lomeli, M., Nalmpantis, C., Pasunuru, R., Raileanu,
  R., Rozi{\`e}re, B., Schick, T., Dwivedi-Yu, J., Celikyilmaz, A., et~al.
\newblock Augmented language models: a survey.
\newblock \emph{arXiv preprint arXiv:2302.07842}, 2023.

\bibitem[Oladimeji \& Sean(2023)Oladimeji and Sean]{oladimeji2023solarwinds}
Oladimeji, S. and Sean, K.
\newblock Solarwinds hack explained: Everything you need to know.
\newblock 2023.
\newblock URL
  \url{https://www.techtarget.com/whatis/feature/SolarWinds-hack-explained-Everything-you-need-to-know}.

\bibitem[OpenAI(2023)]{new2023openai}
OpenAI.
\newblock New models and developer products announced at devday, 2023.
\newblock URL
  \url{https://openai.com/blog/new-models-and-developer-products-announced-at-devday}.

\bibitem[Pa~Pa et~al.(2023)Pa~Pa, Tanizaki, Kou, Van~Eeten, Yoshioka, and
  Matsumoto]{pa2023attacker}
Pa~Pa, Y.~M., Tanizaki, S., Kou, T., Van~Eeten, M., Yoshioka, K., and
  Matsumoto, T.
\newblock An attacker’s dream? exploring the capabilities of chatgpt for
  developing malware.
\newblock In \emph{Proceedings of the 16th Cyber Security Experimentation and
  Test Workshop}, pp.\  10--18, 2023.

\bibitem[playwright(2023)]{playwright2023}
playwright.
\newblock Playwright: Fast and reliable end-to-end testing for modern web apps,
  2023.
\newblock URL \url{https://playwright.dev/}.

\bibitem[Qi et~al.(2023)Qi, Zeng, Xie, Chen, Jia, Mittal, and
  Henderson]{qi2023fine}
Qi, X., Zeng, Y., Xie, T., Chen, P.-Y., Jia, R., Mittal, P., and Henderson, P.
\newblock Fine-tuning aligned language models compromises safety, even when
  users do not intend to!
\newblock \emph{arXiv preprint arXiv:2310.03693}, 2023.

\bibitem[Regina et~al.(2020)Regina, Meyer, and Goutal]{regina2020text}
Regina, M., Meyer, M., and Goutal, S.
\newblock Text data augmentation: Towards better detection of spear-phishing
  emails.
\newblock \emph{arXiv preprint arXiv:2007.02033}, 2020.

\bibitem[Research(2024)]{noushermes2024}
Research, N.
\newblock Nous hermes 2 - yi-34b, 2024.
\newblock URL \url{https://huggingface.co/NousResearch/Nous-Hermes-2-Yi-34B}.

\bibitem[Satter \& Bing(2023)Satter and Bing]{satter2023us}
Satter, R. and Bing, C.
\newblock Us officials seize extortion websites; ransomware hackers vow more
  attacks.
\newblock 2023.
\newblock URL
  \url{https://www.reuters.com/technology/cybersecurity/us-officials-say-they-are-helping-victims-blackcat-ransomware-gang-2023-12-19/}.

\bibitem[Schick et~al.(2023)Schick, Dwivedi-Yu, Dess{\`\i}, Raileanu, Lomeli,
  Zettlemoyer, Cancedda, and Scialom]{schick2023toolformer}
Schick, T., Dwivedi-Yu, J., Dess{\`\i}, R., Raileanu, R., Lomeli, M.,
  Zettlemoyer, L., Cancedda, N., and Scialom, T.
\newblock Toolformer: Language models can teach themselves to use tools.
\newblock \emph{arXiv preprint arXiv:2302.04761}, 2023.

\bibitem[Seymour \& Tully(2018)Seymour and Tully]{seymour2018generative}
Seymour, J. and Tully, P.
\newblock Generative models for spear phishing posts on social media.
\newblock \emph{arXiv preprint arXiv:1802.05196}, 2018.

\bibitem[Shinn et~al.(2023)Shinn, Cassano, Gopinath, Narasimhan, and
  Yao]{shinn2023reflexion}
Shinn, N., Cassano, F., Gopinath, A., Narasimhan, K.~R., and Yao, S.
\newblock Reflexion: Language agents with verbal reinforcement learning.
\newblock In \emph{Thirty-seventh Conference on Neural Information Processing
  Systems}, 2023.

\bibitem[Sikorski \& Honig(2012)Sikorski and Honig]{sikorski2012practical}
Sikorski, M. and Honig, A.
\newblock \emph{Practical malware analysis: the hands-on guide to dissecting
  malicious software}.
\newblock no starch press, 2012.

\bibitem[Teknium(2024)]{openhermes2024}
Teknium.
\newblock Openhermes 2.5 - mistral 7b, 2024.
\newblock URL \url{https://huggingface.co/teknium/OpenHermes-2.5-Mistral-7B}.

\bibitem[Touvron et~al.(2023)Touvron, Martin, Stone, Albert, Almahairi, Babaei,
  Bashlykov, Batra, Bhargava, Bhosale, et~al.]{touvron2023llama}
Touvron, H., Martin, L., Stone, K., Albert, P., Almahairi, A., Babaei, Y.,
  Bashlykov, N., Batra, S., Bhargava, P., Bhosale, S., et~al.
\newblock Llama 2: Open foundation and fine-tuned chat models.
\newblock \emph{arXiv preprint arXiv:2307.09288}, 2023.

\bibitem[Varshney(2023)]{varshney2023introduction}
Varshney, T.
\newblock Introduction to llm agents.
\newblock 2023.
\newblock URL
  \url{https://developer.nvidia.com/blog/introduction-to-llm-agents/}.

\bibitem[Wang et~al.(2023{\natexlab{a}})Wang, Cheng, Zhan, Li, Song, and
  Liu]{wang2023openchat}
Wang, G., Cheng, S., Zhan, X., Li, X., Song, S., and Liu, Y.
\newblock Openchat: Advancing open-source language models with mixed-quality
  data.
\newblock \emph{arXiv preprint arXiv:2309.11235}, 2023{\natexlab{a}}.

\bibitem[Wang et~al.(2023{\natexlab{b}})Wang, Wang, Liu, Chen, Yuan, Peng, and
  Ji]{wang2023mint}
Wang, X., Wang, Z., Liu, J., Chen, Y., Yuan, L., Peng, H., and Ji, H.
\newblock Mint: Evaluating llms in multi-turn interaction with tools and
  language feedback.
\newblock \emph{arXiv preprint arXiv:2309.10691}, 2023{\natexlab{b}}.

\bibitem[Wei et~al.(2022{\natexlab{a}})Wei, Tay, Bommasani, Raffel, Zoph,
  Borgeaud, Yogatama, Bosma, Zhou, Metzler, et~al.]{wei2022emergent}
Wei, J., Tay, Y., Bommasani, R., Raffel, C., Zoph, B., Borgeaud, S., Yogatama,
  D., Bosma, M., Zhou, D., Metzler, D., et~al.
\newblock Emergent abilities of large language models.
\newblock \emph{arXiv preprint arXiv:2206.07682}, 2022{\natexlab{a}}.

\bibitem[Wei et~al.(2022{\natexlab{b}})Wei, Wang, Schuurmans, Bosma, Xia, Chi,
  Le, Zhou, et~al.]{wei2022chain}
Wei, J., Wang, X., Schuurmans, D., Bosma, M., Xia, F., Chi, E., Le, Q.~V.,
  Zhou, D., et~al.
\newblock Chain-of-thought prompting elicits reasoning in large language
  models.
\newblock \emph{Advances in Neural Information Processing Systems},
  35:\penalty0 24824--24837, 2022{\natexlab{b}}.

\bibitem[Weng(2023)]{weng2023llm}
Weng, L.
\newblock Llm powered autonomous agents, 2023.
\newblock URL \url{https://lilianweng.github.io/posts/2023-06-23-agent/}.

\bibitem[Xi et~al.(2023)Xi, Chen, Guo, He, Ding, Hong, Zhang, Wang, Jin, Zhou,
  et~al.]{xi2023rise}
Xi, Z., Chen, W., Guo, X., He, W., Ding, Y., Hong, B., Zhang, M., Wang, J.,
  Jin, S., Zhou, E., et~al.
\newblock The rise and potential of large language model based agents: A
  survey.
\newblock \emph{arXiv preprint arXiv:2309.07864}, 2023.

\bibitem[Yang et~al.(2023)Yang, Wang, Zhang, Petzold, Wang, Zhao, and
  Lin]{yang2023shadow}
Yang, X., Wang, X., Zhang, Q., Petzold, L., Wang, W.~Y., Zhao, X., and Lin, D.
\newblock Shadow alignment: The ease of subverting safely-aligned language
  models.
\newblock \emph{arXiv preprint arXiv:2310.02949}, 2023.

\bibitem[Yao et~al.(2022)Yao, Zhao, Yu, Du, Shafran, Narasimhan, and
  Cao]{yao2022react}
Yao, S., Zhao, J., Yu, D., Du, N., Shafran, I., Narasimhan, K., and Cao, Y.
\newblock React: Synergizing reasoning and acting in language models.
\newblock \emph{arXiv preprint arXiv:2210.03629}, 2022.

\bibitem[Zhan et~al.(2023)Zhan, Fang, Bindu, Gupta, Hashimoto, and
  Kang]{zhan2023removing}
Zhan, Q., Fang, R., Bindu, R., Gupta, A., Hashimoto, T., and Kang, D.
\newblock Removing rlhf protections in gpt-4 via fine-tuning.
\newblock \emph{arXiv preprint arXiv:2311.05553}, 2023.

\bibitem[Zheng et~al.(2023)Zheng, Chiang, Sheng, Zhuang, Wu, Zhuang, Lin, Li,
  Li, Xing, Zhang, Gonzalez, and Stoica]{zheng2023judging}
Zheng, L., Chiang, W.-L., Sheng, Y., Zhuang, S., Wu, Z., Zhuang, Y., Lin, Z.,
  Li, Z., Li, D., Xing, E.~P., Zhang, H., Gonzalez, J.~E., and Stoica, I.
\newblock Judging llm-as-a-judge with mt-bench and chatbot arena, 2023.

\bibitem[Zou et~al.(2023)Zou, Wang, Kolter, and Fredrikson]{zou2023universal}
Zou, A., Wang, Z., Kolter, J.~Z., and Fredrikson, M.
\newblock Universal and transferable adversarial attacks on aligned language
  models.
\newblock \emph{arXiv preprint arXiv:2307.15043}, 2023.

\end{thebibliography}
\bibliographystyle{icml2024}




\end{document}